\documentclass{article}
\usepackage{spconf,graphicx,color}
\usepackage{amsmath}
\usepackage[toc,page]{appendix}

\usepackage{cite}
\usepackage{ amssymb }
\usepackage{bm} 
\usepackage{amsthm}

 \newtheorem{lemma}{Lemma}
 
 \theoremstyle{remark}

\newcommand{\LCM}{\mathop{\mathrm{LCM}}\nolimits}

\usepackage{algorithm}
\usepackage{algpseudocode}
\usepackage[caption=false,font=footnotesize]{subfig}
\title{FPS-SFT: A Multi-dimensional Sparse Fourier Transform Based on the Fourier Projection-slice Theorem}
\name{Shaogang Wang,  Vishal M. Patel and Athina Petropulu
\thanks{This paper has been submitted to IEEE ICASSP 2018.}
}
\address{Department of Electrical and Computer Engineering\\
Rutgers, The State University of New Jersey, Piscataway, NJ 08854, USA}
\begin{document}
\ninept
\maketitle
\begin{abstract}
We propose a  multi-dimensional (M-D) sparse Fourier transform inspired by the idea of the Fourier  projection-slice theorem, called FPS-SFT. FPS-SFT extracts 
samples along lines ($1$-dimensional slices from an M-D data cube), which are parameterized by random slopes and offsets. The discrete Fourier transform (DFT) along those lines represents projections of M-D DFT of the M-D data onto those lines. The M-D sinusoids that are contained in the signal can be reconstructed from the DFT along lines with a low sample and computational complexity provided that the signal is sparse in the frequency domain and the lines are appropriately designed.  The performance of FPS-SFT is demonstrated both theoretically and numerically. A sparse image reconstruction application is illustrated, which shows the capability of the FPS-SFT in solving practical problems.
\end{abstract}
\section{Introduction}
\label{sec:intro}
Conventional signal processing methods in radar, sonar, and medical imaging systems usually involve  multi-dimensional discrete Fourier transforms (DFT), which can be implemented by the fast Fourier transform (FFT). The sample and computational complexity of the FFT are $O(N)$ and $O(N \log N)$, respectively, where $N$ is the number of samples in the multi-dimensional sample space. 
%
%
Recently, the sparse Fourier transform (SFT) \cite{hassanieh2012nearly,ghazi2013sample,potts2015sparse,pawar2017ffast} has been proposed, which leverages the sparsity of signals in the frequency domain to reduce the sample and computational cost of the FFT.  Different versions of the SFT  have been investigated for several applications including a fast Global Positioning System (GPS) receiver, wide-band spectrum sensing, radar signal processing, etc.  \cite{hassanieh2012faster, hassanieh2014ghz,wang2017robust,shi2014light,hassanieh2015fast}.

\sloppy Multi-dimensional signal processing requires multi-dimensional SFT algorithms. The $2$-dimensional ($2$-D) SFT algorithm proposed in \cite{ghazi2013sample} achieves  sample complexity $O(K)$ and computational complexity of $O(K \log K)$, which are the lower bounds of the complexities of known SFT algorithms to date\cite{pawar2017ffast}. The reduction of complexity in SFT of \cite{ghazi2013sample} is achieved  by implementing a $2$-D DFT as a series of $1$-dimensional ($1$-D) DFTs, which are applied on a few columns and rows of the input data matrix. 
The SFT of \cite{ghazi2013sample} basically extends the $1$-D SFT algorithm of \cite{hassanieh2012nearly} to two dimensions; such SFT algorithm employs the so-called OFDM-trick to decode the frequencies that are embedded in the phase difference of DFTs of the same signal but with different sample offsets.
However, the SFT of \cite{ghazi2013sample} only applies to the $2$-D cases with equal sample length, $\sqrt{N}$, of the two dimensions; $\sqrt{N}$ is assumed to be a power of $2$. Moreover, to achieve a high success rate of frequency recovery, it assumes that the data is very sparse ($K << \sqrt{N}$) in the $2$-D frequency domain and the frequency locations are distributed uniformly.  However, those assumptions are not always valid in  practical scenarios.  
  
In this work, we propose FPS-SFT, a new SFT algorithm that uses the basic idea of \cite{ghazi2013sample} while avoiding the shortcomings of \cite{ghazi2013sample}, and can be generalized to the  $D$-dimensional ($D$-D), $D \ge 2$ cases.
The FPS-SFT implements a $D$-D DFT via a sequence of $1$-D DFTs, applied on samples of the $D$-D data which are taken along discrete lines; the lines  are parametrized with random slopes and offsets. 
This is different from \cite{ghazi2013sample}, where the lines are restricted along the axis of each dimension, i.e., the rows and the columns. 
The proposed FPS-SFT can be viewed as a low-complexity, Fourier projection-slice approach for signals that are sparse in the frequency domain.
In the Fourier projection-slice theorem \cite{mersereau1974digital}, the Fourier transform of a projection is a slice of the $D$-D Fourier transform along the same line the projection was taken. 
In FPS-SFT, the $1$-D DFT along a line, which is a $1$-D slice of the $D$-D data is the projection of the $D$-D DFT of the $D$-D data to such line. 
While the classic Fourier projection-slice based method reconstructs the frequency domain of the signal using interpolation based on  frequency-domain slices, the FPS-SFT aims to reconstruct the signal based on DFT of  time-domain slices with reduced complexity; this is achieved by leveraging the sparsity of the signal in the frequency domain.  

The connection between SFT algorithms and the Fourier projection-slice theorem is also found in \cite{shi2014light,hassanieh2015fast}, where the SFT algorithms also rely on lines extracted from $D$-D data. The recovery of the frequency locations in those SFT algorithms are based on a voting procedure; specifically, each entry of the DFT along a line is the projection of the $D$-D DFT of the data; the projected DFT values lie in a $D-1$-dimensional hyper-plane, which is orthogonal to the time-domain line. When the entry value of the DFT along a line is significant, each DFT grid in the $D-1$-dimensional hyperplane gains one vote. After applying DFT on a sufficient number of lines with different slopes followed by the voting procedure, the DFT grids with the largest number of votes are recovered as the significant frequencies. When $K$ is moderately large, such method would generate many false frequencies due to that many zero-valued frequency locations also gain large votes stemming from the ambiguity in the voting process. Moreover, the sample and computational complexity of those SFT algorithms do not achieve the lower bounds of the state-of-the-art SFT algorithms\cite{ghazi2013sample,pawar2017ffast}.  

The fundamental difference between the FPS-SFT and the SFT algorithms of \cite{shi2014light,hassanieh2015fast} is that the FPS-SFT is inspired by the low-complexity SFT of \cite{ghazi2013sample}, which essentially utilizes  phase information to recover the significant frequencies in a progressive manner, i.e., each iteration in the FPS-SFT recovers a subset of significant frequencies, whose contributions are removed in  subsequent iterations; this results in a sparser signal. 
 
The advantages of the proposed FPS-SFT are summarized as follows. FPS-SFT applies to data of arbitrary dimensions and sizes, which are sparse in the frequency domain. 
In the $2$-D cases, the FPS-SFT outperforms the SFT of \cite{ghazi2013sample} significantly when the sparsity of the data reduces. The limitation of the SFT of \cite{ghazi2013sample} on $K$-sparse signals with large $K$ and uniformly distributed frequencies essentially stems from the fact that the direction of projection in the DFT domain is restricted to be along rows and columns. By randomizing the direction of projection in the DFT domain, achieved by taking DFT along lines with pseudo-random slopes, the FPS-DFT can accommodate signals that contain less sparse, non-uniformly distributed frequencies. 


\smallskip
\noindent \textbf{Notation:} 
We use lower-case bold letters to denote vectors. $[\cdot]^T$ denotes the transpose of a vector. The $N$-modulo operation is denoted by $[\cdot]_N$. $[S]$ refers to the integer set of $\{0, ... , S-1  \}$. 
The cardinality of set $\mathbb{S}$ is denoted as $|\mathbb{S}|$. We use $a\perp b$ to denote that $a$ and $b$ are co-prime. The DFT of signal $x$ is denoted by $\hat{x}$.

\section{The FPS-SFT Algorithm} \label{sec:algorithm}

We consider the following $2$-D  signal model, which is a superposition of $K$ $2$-D complex sinusoids, i.e., 
\begin{equation} \label{eq:sigModel2}
x(\mathbf{n}) \triangleq \sum_{(a, \bm{\omega}) \in \mathbb{S}_2} a e^{j \mathbf{n}^T \bm{\omega} }, 
\end{equation}
where $\mathbf{n} \triangleq [n_0, n_1]^T \in \mathcal{X}_2 \triangleq [N_0] \times [N_1]$, with $N_0,N_1$ denoting the sample length of the two dimensions, respectively. $(a, \bm{\omega})$ represents a $2$-D sinusoid whose amplitude is $a$ with $a \in \mathbb{C}, a \neq 0$ and frequency is $\bm{\omega} \triangleq [\omega_0, \omega_1]^T$ with $\omega_k =\frac{2 \pi}{N_k} m_k,  m_k \in [N_k], k \in \{0,1\}$. The set $\mathbb{S}_2$ with $|\mathbb{S}_2|=K$ includes all the sinusoids. We assume that the signal is sparse in the frequency domain, i.e., $K << N \triangleq N_0 N_1$. 
The problem we address is the recovery of $\mathbb{S}_2$ from samples of $x(\mathbf{n})$. The generalization to the higher dimension, i.e., $D$-D cases with $D > 2$ is straightforward. 

\subsection{The SFT algorithm of \cite{ghazi2013sample}} \label{sec:GHIKPS_SFT}
According to \cite{ghazi2013sample},   in order to recover the frequency set $\mathbb{S}_2$, $1$-D DFTs are applied on a subset of columns and rows of the data.  
%
%
The $N_0$-point DFT of the $i_{\rm{th}}, i \in [N_1]$ column of the data equals 
\begin{equation*}
\begin{split}
\hat{c}_i (m) &\triangleq \frac{1}{N_0} \sum_{l\in [N_0]} x(l,i) e^{-j \frac{2\pi}{N_0} ml}  \\
&= \frac{1}{N_0}  \sum_{(a, \bm{\omega}) \in \mathbb{S}_2} \sum_{l\in [N_0]} a e^{j \frac{2\pi}{N_1}  m_1 i} e^{j  \frac{2\pi}{N_0} l (m_0 - m)}, \; m \in [N_0].
\end{split}
\end{equation*}
For a fixed $m$, $\hat{c}_i (m)$ is the summation of  modulated amplitudes of the $2$-D sinusoids, 
$(a, [2\pi m_0/N_0, 2\pi m_1/N_1]^T) \in \mathbb{S}_2$, whose frequencies lie on line
\begin{equation} \label{eq:column_line}
m_0 - m = 0, m_0 \in [N_1],
\end{equation}
which is a row in the $N_0\times N_1$-point DFT of (\ref{eq:sigModel2}), i.e., $\hat{x}(m_0,m_1), [m_0,m_1]^T \in \mathcal{X}_2$. Thus, $\hat{c}_i (m), m\in [N_0]$, the DFT along a column, can be viewed as the projection of $\hat{x}(m_0,m_1)$ on that column. 
Similarly, the $N_1$-point DFT applied on a row of (\ref{eq:sigModel2}) are projections of columns of $\hat{x}(m_0,m_1)$ on that row.
Since the signal is sparse in the frequency domain, if $|\hat{c}_i (m)| \neq 0$, with high probability, there is only one significant frequency laying on line (\ref{eq:column_line}); in such case, we call the frequency bin $m$ to be `$1$-sparse',  and  $\hat{c}_i (m)$ is reduced to be $\hat{c}_i (m) = \hat{c}_i (m_0) =  a e^{j \frac{2\pi}{N_1} m_1 i}$. 
The amplitude, $a$, can be determined by the $m_0$-th entry of the DFT of the $0$-th column, i.e., $a = \hat{c}_0 (m_0)$, and the other frequency component, $m_1$, is `coded' in the phased difference between the $m_0$-th entries of the DFTs of the $0$-th and the $1$-st columns, which can be decoded by $m_1 = \phi \left( {\hat{c}_1 (m_0)}/{ \hat{c}_0 (m_0)} \right) \frac{N_1}{2 \pi} $, where $\phi(x)$ is the phase of $x$. Note that the $1$-sparsity of the $m_{\rm{th}}$ bin  can be effectively tested by comparing $|\hat{c}_0 (m)|$ and $|\hat{c}_1 (m)|$, i.e., $\hat{c}_i (m)$ is $1$-sparse almost for sure when $|\hat{c}_0 (m)| = |\hat{c}_1 (m)|$. Such frequency decoding technique is referred to as  OFDM-trick \cite{hassanieh2012nearly}. The decoded frequencies are removed from the signal, so that the following processing can be  applied on a sparser signal, which is likely to generate more $1$-sparse bins in the subsequent processing.

A frequency bin that is not $1$-sparse in column processing might be $1$-sparse in row processing. Also, the removal of frequencies in the column (row) processing may cause  bins in the row (column) processing to be $1$-sparse, the SFT of \cite{ghazi2013sample} runs iteratively and alternatively between columns and rows. The algorithm stops after a finite number of iterations. 

The SFT of \cite{ghazi2013sample} succeeds with high probability only when the frequencies are  very sparse; this is due to the `deadlock' structures that exist in the distribution of   frequency locations. In a deadlock case, neither a column nor a row DFT contains a $1$-sparse bin. In fact, in many applications, the signal frequency exhibits a block sparsity pattern \cite{eldar2010block}, i.e., the significant frequencies are clustered. In those cases, even when the signal is very sparse, deadlocks are inevitable. 

\subsection{FPS-SFT} \label{sec:DL_SFT_overview}
\begin{figure}[!htp]
	\begin{center}
	\includegraphics[scale=0.5]{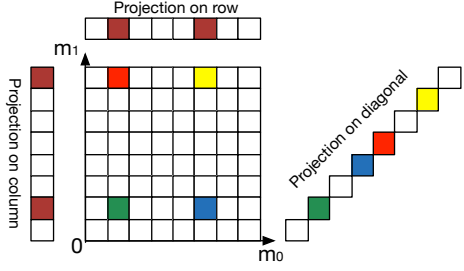} 
\vskip -20pt	\caption{Demonstration of $2$-D frequencies projecting onto $1$-D. The projection onto a column or a row causes collisions, while the projection onto the diagonal creates $1$-sparse bins. The colored blocks mark significant frequencies.}
	\label{fig:freqProjection}
	\end{center}
\end{figure}

The SFT of \cite{ghazi2013sample} reduces a $2$-D DFT into $1$-D DFTs of the columns and rows of the input data matrix. The columns and the rows can be viewed as discrete lines of the input data matrix with slopes $\infty$ and $0$, respectively. In this section, by proposing FPS-SFT, we reduce the $2$-D DFT into $1$-D DFTs of the data along discrete lines with random slopes and offsets. 
The SFT of \cite{ghazi2013sample} resolves $2$-D frequencies that are projected to $1$-sparse bins of the column and row DFTs, and a deadlock arises when such projections cannot create any $1$-sparse bins. In FPS-SFT, by employing lines with random slopes, the directions of  projection are also random, which offers a high probability of creating more $1$-sparse bins and resolving the deadlocks encountered by SFT of \cite{ghazi2013sample}. This can be illustrated in Fig. \ref{fig:freqProjection}, where the $4$ $2$-D frequencies in the $8 \times 8$-point DFT domain form a deadlock, as neither a row DFT nor a column DFT creates a $1$-sparse bin. However, the DFT along the diagonal, corresponding to the projection of the $2$-D DFT of data onto the diagonal, produces $4$ $1$-sparse bins, which solves the deadlock.

FPS-SFT is an iterative algorithm; each iteration returns a subset of recovered $2$-D frequencies. After $T$ iterations, the FPS-SFT returns a frequency set, $\hat{\mathbb{S}}_2$, which is an estimate of $\mathbb{S}_2$ (see (\ref{eq:sigModel2})). The frequencies recovered in previous iterations are passed to the next iteration, and their contributions are removed from the signal in order to create a sparser signal. 

Within each iteration of FPS-SFT, the signal of (\ref{eq:sigModel2}) is sampled along a line with slope $\alpha_1/\alpha_0$ starting at point $(\tau_0,\tau_1)$, with  $\bm{\alpha}, \bm{\tau} \in \mathcal{X}_2$, where $\bm{\alpha} \triangleq [\alpha_0, \alpha_1]^T, \bm{\tau} \triangleq [\tau_0,\tau_1]^T$. The sampled signal can be expressed as  
\begin{equation} \label{eq:sample_slice}
\begin{split}
&s(\bm{\alpha},\bm{\tau},l) \triangleq x([\alpha_0 l+\tau_0]_{N_0}, [\alpha_1 l+\tau_1]_{N_1})\\  &= \sum_{(a,\bm{\omega})\in \mathbb{S}_2} a e^{j 2\pi  \left( \frac {m_0  [\alpha_0 l+\tau_0]_{N_0}}{N_0} + \frac{m_1  [\alpha_1 l+\tau_1]_{N_1}}{N_1} \right)},   l\in [L].
\end{split}
\end{equation}
Note that such line can wrap around within $x(n_0, n_1)$, and the sampling points along the line are always on the grid of $x(n_0, n_1)$ due to the choice of  $\bm{\alpha}, \bm{\tau}$.

On taking an $L$-point DFT on (\ref{eq:sample_slice}), we get
\begin{equation}  \label{eq:hs}
\begin{split}
&\hat{s}(\bm{\alpha},\bm{\tau},m) \triangleq \frac{1}{L} \sum_{l \in [L]}  s(\bm{\alpha},\bm{\tau}, l) e^{-j 2\pi \frac{l m}{ L}} \\
 &= \frac{1}{L} \sum_{(a,\bm{\omega})\in \mathbb{S}_2} a e^{j 2\pi \left(\frac{m_0 \tau_0}{N_0} + \frac{m_1 \tau_1}{N_1} \right)} \sum_{l \in [L]} e^{j2\pi l \left(\frac{m_0 \alpha_0}{N_0} + \frac{m_1 \alpha_1}{N_1} - \frac{m}{L}\right)},
\end{split}
\end{equation}
where $m \in [L]$.

Let us  assume that the line parameters are designed such that the  orthogonality condition for frequency projection are satisfied (see Lemma \ref{le:length} for details), i.e., for $m \in [L], [m_0,m_1]^T \in \mathcal{X}_2$, 
\begin{equation} \label{eq:orthogonal}
\begin{split}
& \hat{f}(m) \triangleq \frac{1}{L} \sum_{l \in [L]} e^{j2\pi l \left(\frac{m_0 \alpha_0}{N_0} + \frac{m_1 \alpha_1}{N_1} - \frac{m}{L}\right)} \in \{0,1\},\\
\end{split}
\end{equation}
then if
\begin{equation} \label{eq:lines}
\left[\frac{m_0 \alpha_0}{N_0} + \frac{m_1 \alpha_1}{N_1} - \frac{m}{L}\right]_1 = 0, [m_0,m_1]^T \in \mathcal{X}_2,
\end{equation}
the $m_{\rm{th}}$ entry of (\ref{eq:hs}) can be simplified as
\begin{equation} \label{eq:freqSum}
\hat{s}(\bm{\alpha},\bm{\tau},m) = \sum_{(a,\bm{\omega}) \in \mathbb{S}_2} a e^{j 2\pi \left(\frac{m_0 \tau_0}{N_0} + \frac{m_1 \tau_1}{N_1} \right)}.
\end{equation} 

Eqs. (\ref{eq:freqSum}) and (\ref{eq:lines}) state that each entry of the $L$-point DFT of the data located along a line with slope $\alpha_1/\alpha_0$ represents a projection of the $2$-D DFT values locating along the line of (\ref{eq:lines}), which is orthogonal to the time domain line (\ref{eq:sample_slice}).  This is closely related to the \emph{Fourier projection-slice theorem}\cite{mersereau1974digital}. The Fourier projection-slice theorem states that the Fourier transform of a projection is a slice of the Fourier transform of the projected object. While the classical projection is in the time domain and the corresponding slice is in the frequency domain, in the FPS-SFT case, the projection is in the DFT domain and the corresponding slice is in the sample (discrete-time) domain. The important difference between the Fourier projection-slice theorem and FPS-SFT is that while the former reconstructs the frequency domain of the signal via interpolation frequency domain slices, which exhibits high complexity,  the latter efficiently recovers the significant frequencies of the signal directly based on the DFT of time-domain $1$-D slices, i.e., samples along random lines.  This is achieved by exploring the sparsity nature of the signal in the frequency domain, which is explained in the following.  

We apply the assumption that the signal is sparse in the frequency domain; specifically, we assume that $|\mathbb{S}_2| = O(L)$. Then,  if $|\hat{s}(\bm{\alpha},\bm{\tau}, m)| \neq 0$, with high probability, the $m_{\rm{th}}$ bin is $1$-sparse, and it holds that $\hat{s}(\bm{\alpha},\bm{\tau}, m) = a e^{j 2\pi \left(\frac{m_0 \tau_0}{N_0} + \frac{m_1 \tau_1}{N_1} \right)}, (a, \bm{\omega}) \in \mathbb{S}_2$.
In such case, the $2$-D sinusoid, $(a, \bm{\omega})$, can be `decoded' by three lines of the same slope but different offsets. The offsets for the three lines are designed as $\bm{\tau}, \bm{\tau}_0 \triangleq [[\tau_0+1]_{N_0}, \tau_1]^T, \bm{\tau}_1 \triangleq [\tau_0, [\tau_1+1]_{N_1}]^T$, respectively; such design allows for the frequencies to be decoded independently in each dimension. The sinusoid corresponding to the $1$-sparse bin, $m$, can be decoded as
\begin{equation} \label{eq:decoding}
\begin{split}
&m_0 = \left[ \frac{N_0}{2 \pi} \phi\left( \frac{\hat{s}(\bm{\alpha},\bm{\tau}_0,m) }{\hat{s}(\bm{\alpha},\bm{\tau},m) } \right)  \right]_{N_0}, \\
&m_1 = \left[ \frac{N_1}{2 \pi} \phi\left(\frac{\hat{s}(\bm{\alpha},\bm{\tau}_1,m)}{\hat{s}(\bm{\alpha},\bm{\tau},m)}    \right) \right]_{N_1},\\
&a = \hat{s}(\bm{\alpha},\bm{\tau},m) e^{-j2\pi(m_0 \tau_0/N_0 + m_1 \tau_1/N_1)}.
\end{split}
\end{equation}
To recover all the sinusoids in $\mathbb{S}_2$ efficiently,  each iteration of FPS-SFT adopts a random choice of line slope (see Lemma \ref{le:slope}) and offset. Furthermore, the contribution of the recovered sinusoids 
in the previous iterations is removed via a \emph{construction-subtraction} approach to creating a  sparser signal in the future iterations. Specifically, assuming that for current iteration, the line slope and offset parameters are selected as $\bm{\alpha},\bm{\tau}$, respectively,   the recovered sinusoids are projected into $L$ frequency bins to construct the DFT along the line, $\hat{s}_r(\bm{\alpha},\bm{\tau}, m) \triangleq \sum_{(a,\bm{\omega}) \in \mathcal{I}_{m}} a e^{j 2\pi \left(\frac{m_0 \tau_0}{N_0} + \frac{m_1 \tau_1}{N_1} \right)}$, $m \in [L]$,  
where $\mathcal{I}_{m}, m \in [L]$ represent the subsets of the recovered sinusoids that are related to the constructed DFT along line via projection, i.e.,  $\mathcal{I}_{m} \triangleq \{(a,\bm{\omega}) : [\frac{m_0 \alpha_0}{N_0} + \frac{m_1 \alpha_1}{N_1} - \frac{m}{L}]_1 = 0, [m_0,m_1]^T \in \mathcal{X}_2 \}, m \in [L]$.

\sloppy Next, the $L$-point inverse DFT (IDFT) is applied on $\hat{s}_r (\bm{\alpha},\bm{\tau},m), m \in [L]$, from which the line, ${s}_r (\bm{\alpha},\bm{\tau},l), l \in [L]$ due to the previously recovered sinusoids are constructed. Subsequently, those constructed line samples are subtracted from the signal samples of the current iteration. Since the contribution of the recovered sinusoids is removed, the signal appears sparser and thus the recovery of the remaining sinusoids is easier in the future iterations.

\subsection{Analysis of FPS-SFT}
In this section we provide some lemmas on the design of the lines used in FPS-SFT. Lemma \ref{le:length} shows the design of the line length to guarantee orthogonality of projection.  Lemma \ref{le:slope} provides candidates of line slopes such that, each bin of the DFT along the line corresponds to the same number of frequencies projected to such bin.  
The uniformity of the projection is likely to create more $1$-sparse bins in the DFT of the lines. The proofs of the lemmas can be found in the Appendices. 

\begin{lemma} \label{le:length}
\textbf{(Line Length):}
Let $L$ be the least common multiple (LCM) of $N_0,N_1$, and $s(\bm{\alpha}, \bm{\tau},l) = x ([\alpha_0 l+\tau_0]_{N_0},[\alpha_1 l+\tau_1]_{N_1})$ with $l \in [L], \bm{\alpha} \triangleq [\alpha_0, \alpha_1]^T, \bm{\tau}\triangleq [\tau_0, \tau_1]^T \in \mathcal{X}_2$ be a discrete line extracted from the signal model of (\ref{eq:sigModel2}).
Then each entry of the $L$-point DFT of  $s(\bm{\alpha}, \bm{\tau},l)$, i.e., $\hat{s}(\bm{\alpha}, \bm{\tau},m), m \in [L]$ is the orthogonal projection of DFT values of the $N_0 \times N_1$-point DFT of (\ref{eq:sigModel2}), whose frequencies locate on the discrete line of $[\frac{m_0}{N_0}   \alpha_0  + \frac{m_1}{N_1}   \alpha_1  - \frac{m }{L}]_1 = 0, [m_0,m_1]^T \in \mathcal{X}_2$. Moreover, $L$ is the minimum length of a line to allow orthogonal projection of DFT values of any frequency location $[m_0, m_1]^T \in \mathcal{X}_2$ with arbitrary choice of $\bm{\alpha} \in \mathcal{X}_2$.     
\end{lemma}


\begin{lemma} \label{le:slope}
\textbf{(Line Slope):}
Let $s(\bm{\alpha}, \bm{\tau},l) = x([\alpha_0 l+\tau_0]_{N_0},[\alpha_1 l+\tau_1]_{N_1})$ with $l \in [L], L=\LCM(N_0,N_1), \bm{\alpha}\triangleq [\alpha_0, \alpha_1]^T  \in \mathcal{A} \subset \mathcal{X}_2, \bm{\tau} \triangleq [\tau_0, \tau_1]^T \in \mathcal{X}_2$ be a discrete line extracted from the signal model of (\ref{eq:sigModel2}), where $\mathcal{A} \triangleq \{\bm{\alpha} : \bm{\alpha} \in \mathcal{X}_2, \alpha_0 \perp \alpha_1, \alpha_0 \perp c_1, \alpha_1 \perp c_0\}$ with $c_0 = L/N_0, c_1 = L/N_1$.
Let $\hat{s} (\bm{\alpha}, \bm{\tau},m), m \in [L]$   be the $L$-point DFT of $s(\bm{\alpha}, \bm{\tau},l), l \in [L]$. Then each entry of $\hat{s} (\bm{\alpha}, \bm{\tau},m), m \in [L]$ is the projection of DFT values located at $N/L$ different frequency locations in $\mathcal{X}_2$, i.e., $|\mathcal{P}_{m}| = N/L$, where $\mathcal{P}_{m} \triangleq \{[m_0,m_1]^T: [\frac{m_0}{N_0}   \alpha_0  + \frac{m_1}{N_1}   \alpha_1  - \frac{m }{L}]_1 = 0,[m_0,m_1]^T \in \mathcal{X}_2 \}$. Moreover, $\mathcal{P}_{m} \cap \mathcal{P}_{m'} = \emptyset$ for $m \neq m', m, m' \in [L]$. Thus, the DFT values of $N$ frequency locations in $\mathcal{X}_2$ are uniformly projected into the $L$ frequency bins of $\hat{s} (\bm{\alpha}, \bm{\tau},m), m \in [L]$.  
\end{lemma}



\noindent \textbf{Complexity analysis:} 
The FPS-SFT executes $T$ iterations; in the $2$-D case, the samples used in each iteration is $3L$ since $3$ $L$-length lines, with $L=\LCM(N_0,N_1)$ are extracted in order to decode the two frequency components of a $2$-D sinusoid (see (\ref{eq:decoding})).  
Hence, the sample complexity of FPS-SFT is $O(3T L) = O(L)$.
The core processing of FPS-SFT is the $L$-point $1$-D DFT, which can be implemented by the FFT with the computational complexity of $O(L \log L)$. The $L$-point IDFT in the construction-subtraction procedure can also be implemented by the FFT. In addition to the FFT, each iteration needs to evaluate $O(K)$ frequencies. Hence the computational complexity of FPS-SFT is $O(L \log L+K)$. Assuming that $K = O (L)$, then the sample and computational complexity can be simplified as $O(K)$ and $O(K \log K)$, respectively, which achieves the lower bounds of the  complexity of known state-of-the-art SFT algorithms \cite{ghazi2013sample,pawar2017ffast}. 

\smallskip
\noindent \textbf{Multi-dimensional extension:}
 For the $D$-D case with the data cube size of $N_0 \times N_1 \times \cdots N_{D-1}$, the line length can be set as $L = \LCM(N_0,\cdots,N_{D-1})$; the slope and offset parameters $[\alpha_0, \cdots, \alpha_{D-1}]^T,  [\tau_0, \cdots, \tau_{D-1}]^T$ is randomly taken from $\mathcal{X}_D \triangleq [N_0] \times [N_1] \times \cdots [N_{D-1}]$. Each iteration extracts $D+1$ $L$-length lines with a same random slope but different offsets from the $D$-D data cube. 
The $0$-th line offset is set to be 
$[\tau_0,\cdots, \tau_{D-1}]^T$, while for the 
$i_{\rm{th}}$ line with $1 \le i \le D-1$, the offset for the $i_{\rm{th}}$ dimension is set to be $[\tau_i+1]_{N_i}$. With such offset parameters, the frequencies can be decoded independently for each dimension.

\section{Numerical Results}
\begin{figure}[!htp]
    \centering
    \subfloat[]{{\includegraphics[scale=0.23]{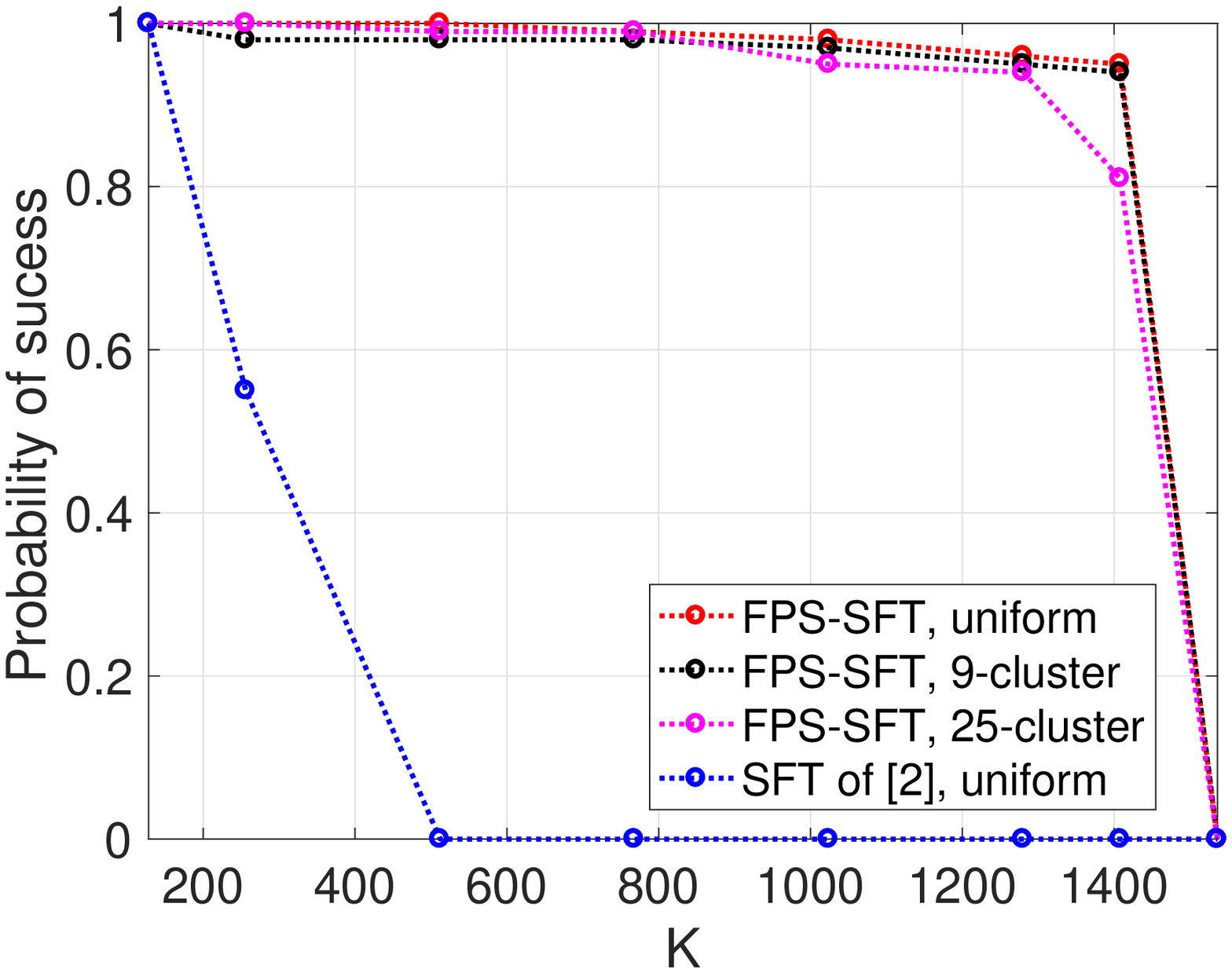} }}%
    \subfloat[]{{\includegraphics[scale=0.23]{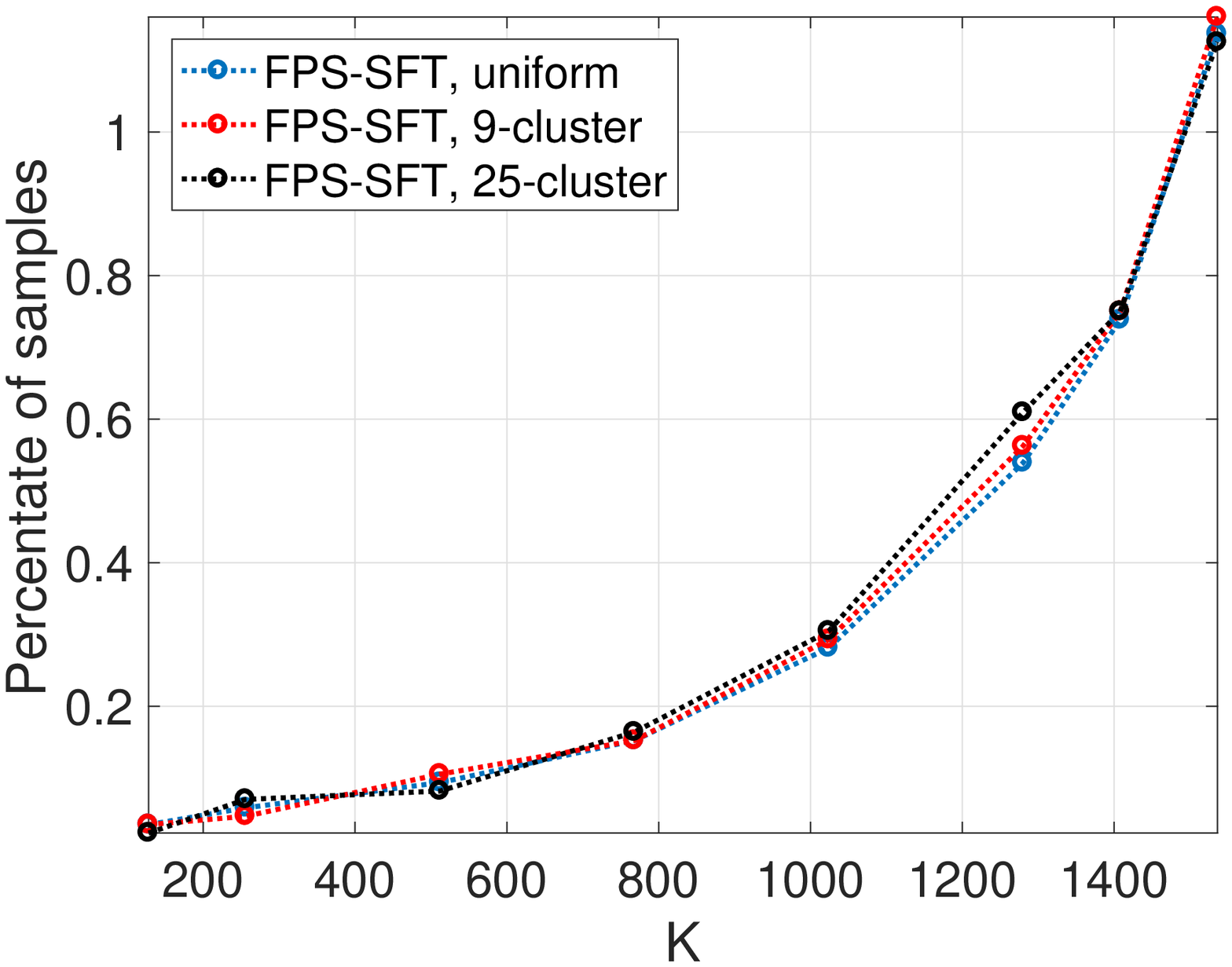} }}%
 \vskip-10pt   \caption{Frequency recovery performance versus sparsity. The results are generated by $100$ iterations of Monte Carlo simulations. (a) Probability of perfect recovery versus number of significant frequencies, $K$. (b) Percentage of samples needed versus $K$.}
    \label{fig:FreqRec}
\end{figure}

\noindent \textbf{Comparison to the SFT of \cite{ghazi2013sample}:}
The length of the two dimensions are set to  $N_0 = N_1 = 256$. We simulate two scenarios, when frequencies are uniformly distributed and when they are clustered. For the clustered case, we consider clusters of $9$ and $25$ frequencies. When $N_0=N_1$, the line length, $L$, of FPS-SFT equals  $N_0$, and each iteration of FPS-SFT uses $3 N_0$ samples. We limit the maximum iterations to $T_{max} = N/(3 L) \approx 85$; this corresponds to roughly  $100\%$ samples of the input data.   
Fig. \ref{fig:FreqRec} (a) shows the probability of perfect recovery versus level of sparsity for FPS-SFT and the SFT of \cite{ghazi2013sample}, respectively. 
When the signal is very sparse, e.g., $K = N_0/2$, the  SFT of \cite{ghazi2013sample} has a high probability for perfect recovery, however, it fails when the sparsity is moderately large, e.g., $K = 2 N_0$. Moreover, the SFT of \cite{ghazi2013sample} only works for the scenario in which frequencies are distributed  uniformly, while it fails when there exists even a single  frequency cluster.  On the contrary, the FPS-SFT applies to signals with a wide range of sparsity levels. For instance, the success rate of FPS-SFT is approximate $96\%$ when $K = 5 N_0$ and the frequency locations are uniformly distributed, while similar performance is observed for the clustered cases considered. In all cases, the success rates drop to $0$ when $K = 6 N_0$, since we set $T_{max} = 85$. To perfectly reconstruct all  frequencies, the FPS-SFT needs to run for roughly $100$ iterations when $K = 6 N_0$. Fig. \ref{fig:FreqRec} (b) shows the percentage of samples used by the FPS-SFT for perfect recovery versus different sparsity level for the uniform and clustered cases. The figure shows that the sparser the signal, the fewer samples are required by the FPS-SFT to recover all the frequencies. For example, when $K = N_0$, only $5.9\%$ of the signal samples are required in the uniform-distributed frequency case or the clustered case. The good performance of FPS-SFT arises because the randomized projections can effectively isolate the frequencies into $1$-sparse bins, even when the signal is less sparse ($K$ is large) and the frequencies are clustered.     


\begin{figure}[htp!]
 \centering 
 \includegraphics[scale=0.2]{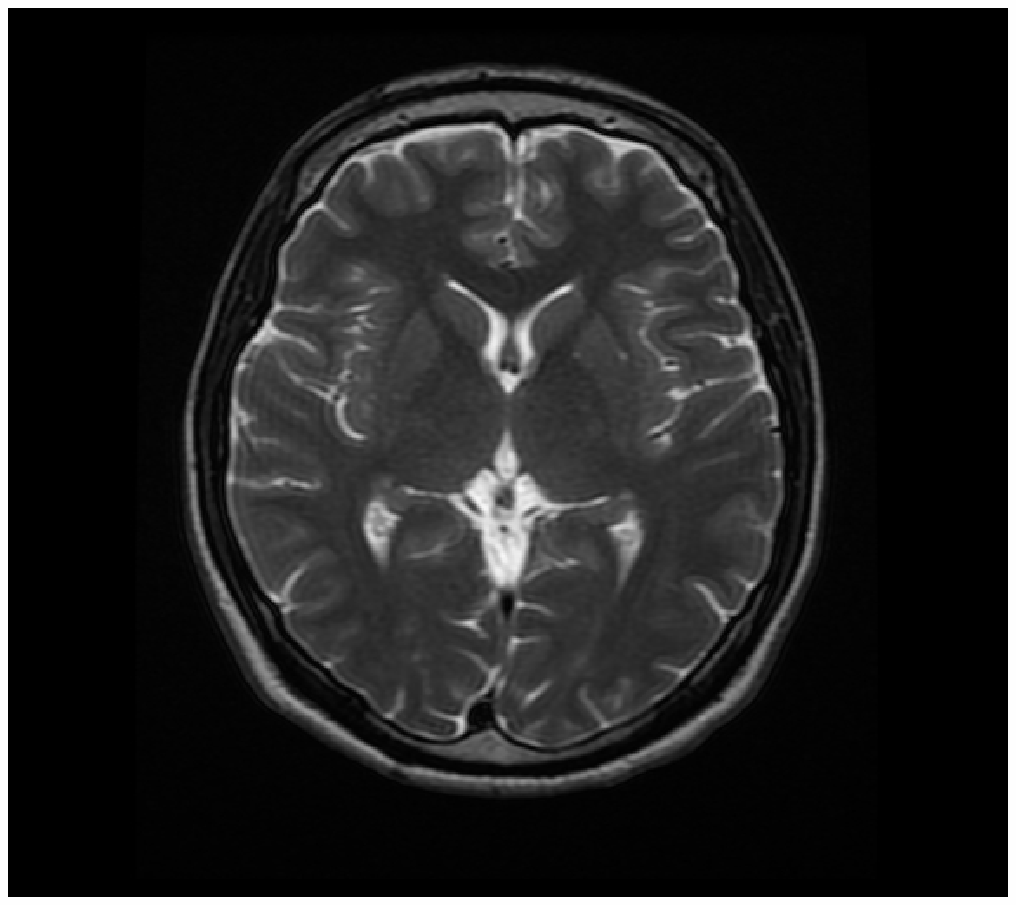} 
 \includegraphics[scale=0.2]{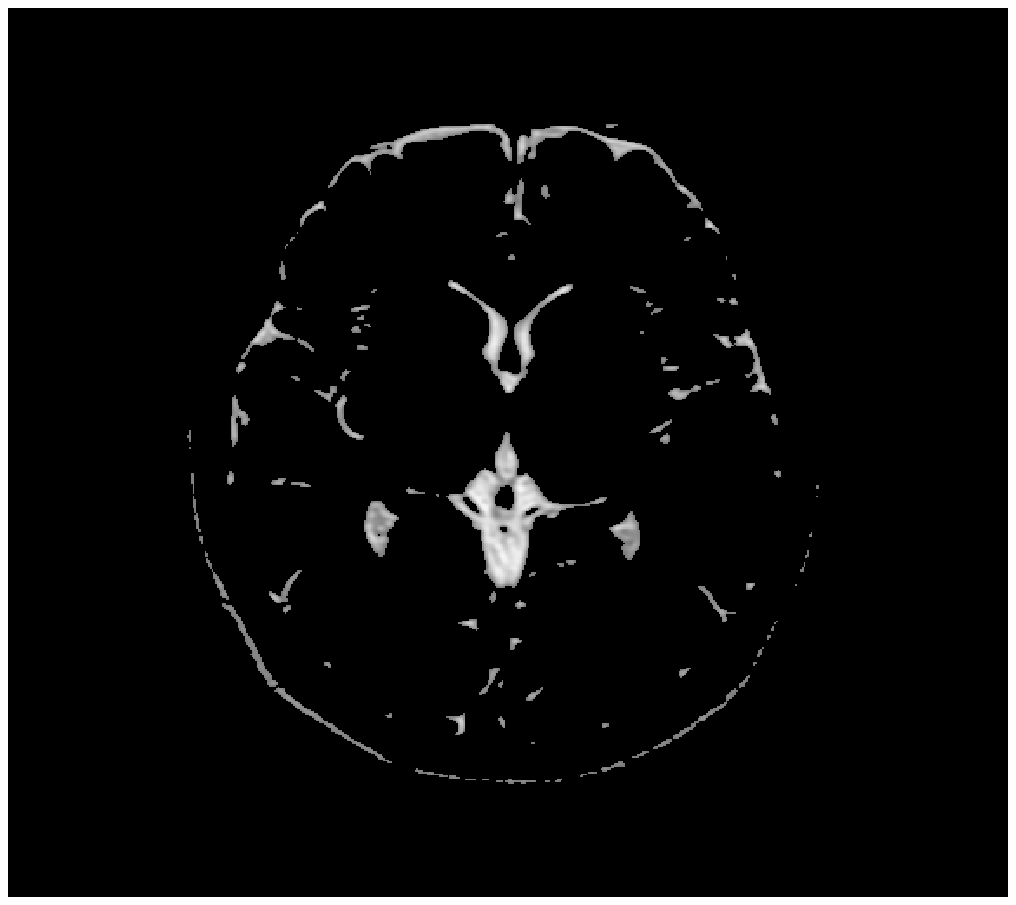}
 \includegraphics[scale=0.2]{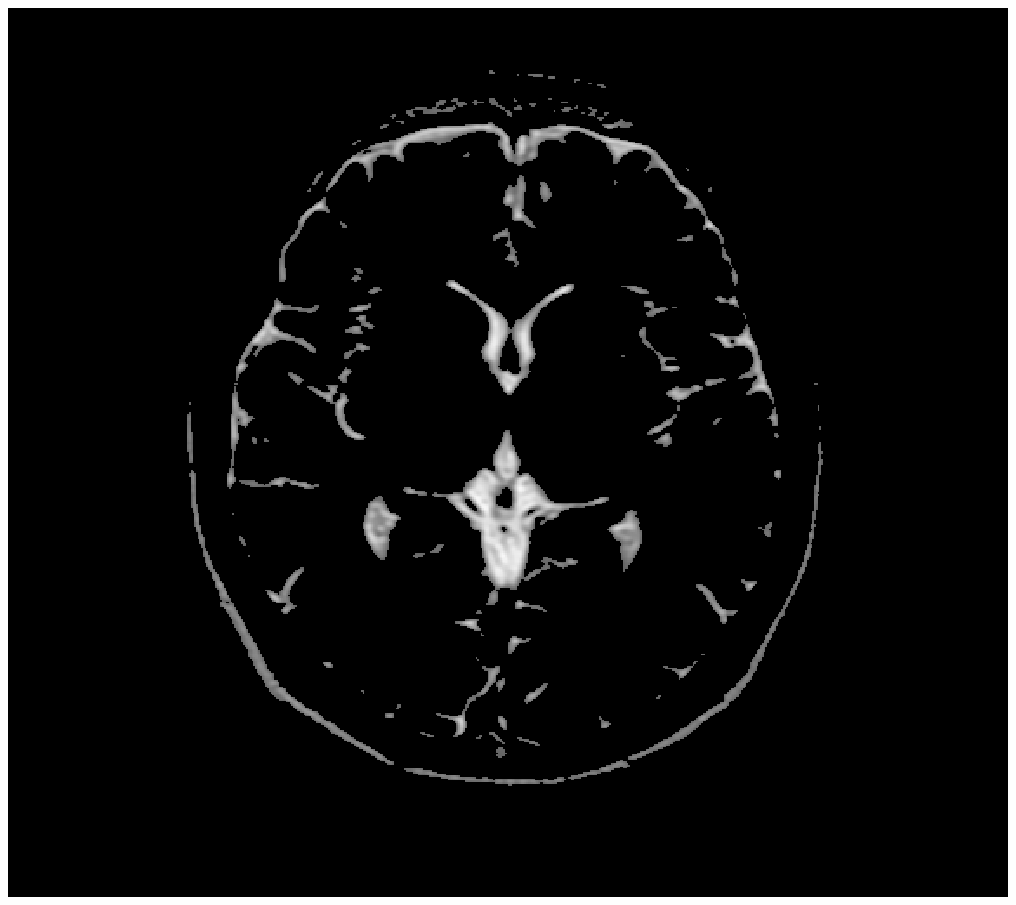} 
 \includegraphics[scale=0.2]{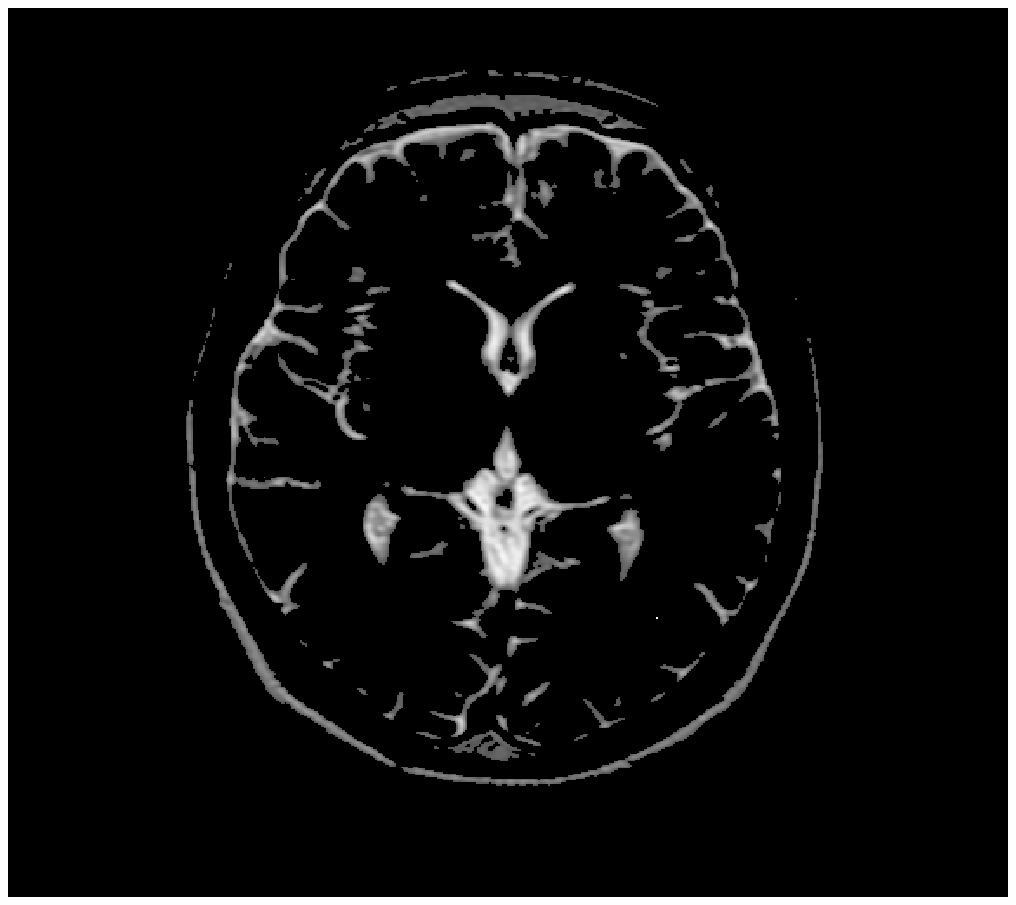}\\
 (a)\hskip50pt(b)\hskip50pt(c)\hskip50pt(d)
   \hskip -15pt \caption{Image reconstruction. (a) Raw  image. (b) $2.9\%$-sparse, $K=8411$. (c) $4.5\%$-sparse, $K=13219$. (d) $6.6\%$-sparse, $K=19506$.}
\label{fig:brain}
\end{figure}

\noindent \textbf{Sparse image reconstruction:}
Due to the duality of the time and frequency, the FPS-SFT is able to reconstruct a signal that is sparse in the time (spatial) domain using the samples in the frequency domain.  Here we demonstrate the ability of FPS-SFT to recover images that are sparse in the pixel domain.
Such sparse image recovery problem arises in the MRI  applications\cite{lustig2008compressed}. In MRI, samples are directly taken from the frequency domain, from which the images reflecting the inner structure of the examined objects are reconstructed. 
Fig. \ref{fig:brain} (a) shows a $512 \times 576$-pixel brain MRI image \cite{lustig2008compressed}. This image was sparsified by   applying thresholding on the original image. Next, we converted the sparsified images into the frequency domain via a $512 \times 576$-point DFT, on which the $2$-D FPS-SFT was applied to reconstruct the images. Figs. \ref{fig:brain} (b), (c) and (d) show that the images with $2.85\%$, $4.48\%$ and $6.61\%$ of non-zero pixels can be perfectly reconstructed by FPS-SFT using $14.0\%$, $23.4\%$, and $70.3\%$ samples in the frequency domain, respectively.


\section{Conclusion} \label{sec:conclusion}
We have proposed the FPS-SFT,  a low-complexity, multi-dimensional SFT algorithm based on the idea of the Fourier projection-slice theorem. Theoretical and numerical results of FPS-SFT have been provided and an application of FPS-SFT on sparse image reconstruction has been demonstrated.

\bibliographystyle{IEEEbib}
\bibliography{SFTbib}

\appendices
\section{Collections of Proofs of Lemmas}
\subsection{Proof of Lemma \ref{le:length}}
\begin{proof}
The orthogonality condition derived in (\ref{eq:orthogonal}) for $[m_0,m_1]^T, [\alpha_0,\alpha_1]^T \in \mathcal{X}_2, m \in [L]$ is equivalent to
\begin{equation} \label{eq:gline}
\left[ \frac{m_0 \alpha_0}{N_0} + \frac{m_1 \alpha_1}{N_1} - \frac{m}{L} \right]_1 = 0,
\end{equation}
which can be rewritten as
\begin{equation} \label{eq:orthogonality}
\left[\frac{L}{N_0} m_0 \alpha_0 + \frac{L}{N_1} m_1 \alpha_1 \right]_L = m.
\end{equation}
It is clear that $L = \LCM(N_0,N_1)$ satisfies the above orthogonality condition, since $L/N_0, L/N_1$ are integers.

Next, we use contradiction to prove that $L = \LCM(N_0,N_1)$ is the smallest line length that allows the orthogonal projection for any $[m_0,m_1]^T, [\alpha_0,\alpha_1]^T \in \mathcal{X}_2$.  

Assume that $L < \LCM(N_0,N_1)$, then, the consequence is that at least either $L/N_0$ or $L/N_1$ is not an integer. Without loss of generality, we assume that $\frac{L}{N_0} \notin \mathbb{Z}$, then the right side of  (\ref{eq:orthogonality}) equals to $[L/N_0]_L \notin [L]$ for $m_0=1,\alpha_0=1, m_1=0$, which is contradictory to the premise that the orthogonality condition holds for any $[m_0,m_1]^T, [\alpha_0,\alpha_1]^T \in \mathcal{X}_2$. Hence $L = \LCM(N_0,N_1)$ is the smallest line length which allows the orthogonal projection of any frequency to a line with arbitrary slope.
\end{proof}

\subsection{Proof of Lemma \ref{le:slope}}
\begin{proof}
This proof is organized as follows. First, by exploring the B\'{e}zout's lemma \cite{rosen1993elementary}, we prove that with the specified line parameters, i.e., $L = \LCM(N_0, N_1), [\alpha_0, \alpha_1]^T \in \mathcal{A}, [\tau_0, \tau_1]^T \in \mathcal{X}_2$, each entry of the DFT along a line, i.e.,  $\hat{s}(\bm{\alpha}, \bm{\tau}, m), m \in [L]$ contains at least the projection of the DFT value from one frequency location $(m_0',m_1')$ in $\mathcal{X}_2$, i.e., $|\mathcal{P}_{m}|>0, m \in [L]$. Next, we prove that $|\mathcal{P}_{m}| \ge N/L$, followed by the proof of $\mathcal{P}_{m} \cap \mathcal{P}_{m'} = \emptyset$ for $m \neq m', m, m' \in [L]$, and finally, we conclude that $|\mathcal{P}_{m}| = N/L$.

Let $\alpha_0' = \alpha_0 c_0, \alpha_1' = \alpha_1 c_1$. 
Since $\alpha_0 \perp \alpha_1, \alpha_0 \perp c_1, \alpha_1 \perp c_0$, and $c_0 \perp c_1$ due to $L = \LCM(N_0, N_1)$, it is obvious that $\alpha_0' \perp \alpha_1'$. According to the 
B\'{e}zout's lemma, there exist $m_0, m_1 \in \mathbb{Z}$, such that 
\begin{equation} \label{eq:bezout}
\alpha_0' m_0 + \alpha_1' m_1 = 1.
\end{equation}
By multiplying $m \in [L]$ to the two sides of (\ref{eq:bezout}), we get
\begin{equation}
\alpha_0' m m_0 + \alpha_1' m m_1 = m ,
\end{equation}
which, using the Euclidean division, can be written as 
\begin{equation} \label{eq:mult_omega}
\alpha'_0 (m_0' + k_0 N_0) + \alpha'_1 (m_1' + k_1 N_1) = m,
\end{equation}
where $m_0' = [m m_0]_{N_0}, m_1'= [m m_1]_{N_1}$; $k_0, k_1 \in \mathbb{Z}$.

Since that
\begin{equation} \label{eq:L}
[\alpha'_0 k_0 N_0 + \alpha'_1 k_1 N_1]_L = [L(\alpha_0 k_0 + \alpha_1 k_1)]_L = 0,
\end{equation}
on taking modulo-$L$ of the two sides of Eq. (\ref{eq:mult_omega}), we have
\begin{equation} \label{eq:solve_line}
[\alpha'_0 m_0' + \alpha'_1 m_1']_L = m,
\end{equation}
which is equivalent to (\ref{eq:gline}). 
It means that there exists a frequency location $[m_0',m_1']^T \in \mathcal{X}_2$, whose DFT value projects to $\hat{s}(\bm{\alpha}, \bm{\tau}, m)$, i.e., $|\mathcal{P}_{m}|>0, m \in [L]$.

Next, let's explore the solution structure of (\ref{eq:solve_line}). It is easy to see that the frequency locations, $[m_0'+k \alpha'_1, m_1' - k \alpha'_0]^T, k\in \mathbb{Z}$, satisfies (\ref{eq:solve_line}), i.e.,
\begin{equation} \label{eq:line_solutions}
[\alpha'_0 (m_0'+k \alpha'_1) + \alpha'_1 (m_1' - k \alpha'_0)]_L = m,
\end{equation}
which can be written as
\begin{equation}
[\alpha'_0 ([m_0'+k \alpha'_1]_{N_0}+k_0 N_0) + \alpha'_1 ([m_1' - k \alpha'_0]_{N_1}+k_1 N_1)]_L = m,
\end{equation}
where $k_0,k_1 \in \mathbb{Z}$. Again, by substituting (\ref{eq:L}), we have
\begin{equation}
[\alpha'_0 [m_0'+k \alpha'_1]_{N_0} + \alpha'_1 [m_1' - k \alpha'_0]_{N_1}]_L = m.
\end{equation}
Hence, the DFT value at frequency locations $\left[[m_0'+k \alpha'_1]_{N_0}, [m_1' - k \alpha'_0]_{N_1}\right]^T \in \mathcal{P}_{m} \subseteq \mathcal{X}_2$, also projects to $\hat{s}(\bm{\alpha}, \bm{\tau}, m)$.

Next, we prove that $|\mathcal{P}_{m}| \ge N/L$. Assume that for $k \neq k'$, there exits two duplicated frequency locations, i.e., $\left[[m_0'+k \alpha'_1]_{N_0}, [m_1' - k \alpha'_0]_{N_1}\right]^T = \left[[m_0'+k' \alpha'_1]_{N_0}, [m_1' - k' \alpha'_0]_{N_1}\right]^T$. It follows that 
\begin{equation}
[k \alpha'_1]_{N_0} = [k' \alpha'_1]_{N_0}, \; [k \alpha'_0]_{N_1} = [k' \alpha'_0]_{N_1},
\end{equation}
which can be rewritten as
\begin{equation} 
k \alpha'_1 = k' \alpha'_1 + k_0 N_0, \; k \alpha'_0 = k' \alpha'_0 + k_1 N_1,
\end{equation}
where $k_0,k_1 \in \mathbb{Z}$. It is easy to conclude that $k_1/k_0 = \alpha_0/\alpha_1$. Hence we have
\begin{equation} 
k \alpha'_1 = k' \alpha'_1 + i \alpha_1 N_0, \; k \alpha'_0 = k' \alpha'_0 + i \alpha_0 N_1, 
\end{equation}
where $i \in \mathbb{Z}, i \neq 0$. Hence 
\begin{equation}
k-k' = i N_0/c_1 = i N/L,
\end{equation}
which means that the frequency location, $\left[[m_0'+k \alpha'_1]_{N_0}, [m_1' - k \alpha'_0]_{N_1}\right]^T$, repeats every $N/L$ points. In another words, there exist at least $N/L$ frequency locations whose DFT values projecting to $\hat{s}(\bm{\alpha}, \bm{\tau}, m)$, i.e., $|\mathcal{P}_{m}|\ge N/L$.

Next, we prove that $\mathcal{P}_{m} \cap \mathcal{P}_{m'} = \emptyset$ for $m \neq m', m, m' \in [L]$. Assume that $[m_0,m_1]^T \in \mathcal{P}_{m} \cap \mathcal{P}_{m'}$,  it can be seen that
\begin{equation}
[\alpha'_0 m_0 + \alpha'_1 m_1]_L = m = m',
\end{equation}
which is contradict with $m \neq m'$. Hence $\mathcal{P}_{m} \cap \mathcal{P}_{m'} = \emptyset$. 

Finally, by combing $\mathcal{P}_{m} \cap \mathcal{P}_{m'} = \emptyset$, $m \in [L]$, $|\mathcal{P}_{m}|\ge N/L$ and $|\mathcal{X}_2| = N$, we can conclude that $|\mathcal{P}_{m}| =  N/L$. This completes the proof.
\end{proof}

\end{document}